\documentclass[11pt,a4paper]{article}

\usepackage[utf8]{inputenc}
\usepackage{textcomp}
\usepackage[margin=2.5cm]{geometry}
\usepackage{amsmath,amssymb}
\usepackage{bm}
\usepackage{graphicx}
\usepackage{xcolor}
\usepackage[font=small,labelfont=bf]{caption}
\usepackage{authblk}
\usepackage[numbers,sort&compress]{natbib}
\usepackage[colorlinks=true,linkcolor=blue,citecolor=blue,urlcolor=blue]{hyperref}

\title{Phase-resolved wide-field CARS microscopy with speckle illumination}

\author[1,*]{Federico Vernuccio}
\author[1,2]{Pascal Berto}
\author[1]{Baptiste Marthy}
\author[1]{Guillaume Baffou}
\author[1]{Sandro Heuke}
\author[3]{Randy Bartels}
\author[4]{Pierre Bon}
\author[1,*]{Herv\'e Rigneault}

\affil[1]{Aix Marseille University, CNRS, Centrale Med, Institut Fresnel, Marseille, France}
\affil[2]{Institut Universitaire de France (IUF), Paris, France}
\affil[3]{Morgridge Institute for Research, Madison, Wisconsin 53715, USA}
\affil[4]{XLIM, CNRS UMR 7252, Universit\'e de Limoges, Limoges, France}
\affil[*]{Corresponding authors: \href{mailto:federico.vernuccio@fresnel.fr}{federico.vernuccio@fresnel.fr}; \href{mailto:herve.rigneault@fresnel.fr}{herve.rigneault@fresnel.fr}}

\date{}

\begin{document}

\maketitle

\begin{abstract}
\textcolor{black}{Coherent anti-Stokes Raman scattering (CARS) microscopy enables label-free chemical imaging of biological samples and materials. Conventionally, CARS is implemented using a point-scanning approach that probes a single vibrational mode at a time. Hyperspectral CARS enhances chemical specificity by sequentially addressing multiple Raman modes. However, the measured CARS intensity is distorted by an undesired non-resonant background, which broadens and shifts the Raman peaks, thereby hindering the interpretability of hyperspectral images. Here, we introduce a phase-sensitive, wide-field hyperspectral CARS microscopy scheme that suppresses the non-resonant background. The proposed approach combines three key components: a high-power picosecond tunable optical parametric amplifier (OPA), speckle illumination, and quantitative phase imaging based on quadriwave lateral shearing interferometry (QLSI). The high-power OPA provides the peak power required for efficient nonlinear excitation. Speckle illumination distributes the optical energy over the objective back pupil and mitigates coherent imaging artifacts. QLSI enables the simultaneous measurement of the amplitude and phase of the CARS field, thereby allowing separation of resonant and non-resonant contributions, without the need for an external reference beam. This unique combination enables practical phase-resolved CARS imaging over a field of view exceeding $60 \times 60~\text{\textmu m}^2$ at a frame rate of $1.4$ Hz. We illustrate the approach by acquiring hyperspectral images of microplastics and liver steatosis across the entire CH-stretching region.}
\end{abstract}

\section{Introduction}

Understanding the chemical composition of complex materials and biological systems is a central challenge in modern optical microscopy. Label-free vibrational imaging techniques provide a powerful means to probe molecular structures without the need for fluorescent markers, thereby preserving the native state of the sample. Among these approaches, coherent Raman scattering (CRS) microscopy has emerged as a particularly attractive modality because it combines chemical specificity with high imaging speed and optical sectioning capabilities \cite{polli_2018,Rigneault2018}. In particular, coherent anti-Stokes Raman scattering (CARS) and stimulated Raman scattering (SRS) microscopy enable the visualization of molecular vibrational signatures with orders-of-magnitude stronger signals than spontaneous Raman scattering, making it well-suited for imaging applications ranging from biology to materials sciences. Conventional CRS microscopy is typically implemented in a point-scanning configuration \cite{Vernuccio2023} where tightly focused and temporally overlapped pump ($\omega_p$) and Stokes ($\omega_s$) beams excite a molecular vibration at frequency $\Omega = \omega_p - \omega_s$. Hyperspectral implementations further enhance chemical specificity by tuning the laser excitation frequencies to sequentially probe multiple vibrational transitions. In SRS, the signal appears as a small intensity modulation, either a loss in the pump beam or a gain in the Stokes beam, typically 4 to 6 orders of magnitude weaker than the incident beams \cite{cheng_stimulated_2021}. Detection, therefore, requires modulation transfer techniques, where one beam is modulated, and the other is detected using a demodulation scheme, commonly via a photodiode coupled to a lock-in amplifier in point-scanning systems. In contrast, CARS arises as a blue-shifted anti-Stokes component $\omega_{as} = 2\omega_p-\omega_s$, which can be spectrally isolated using short-pass filters prior to detection. 
While the measured SRS intensity is directly proportional to the imaginary part of the third-order nonlinear resonant susceptibility and is therefore directly comparable to spontaneous Raman spectra \cite{Hellwarth_77}, the measured CARS intensity scales quadratically with the third-order nonlinear vibrational susceptibility $\chi^{(3)}$ which includes both a complex resonant $\chi^{(3)}_{\text{R}}$ and a non-resonant term $\chi^{(3)}_{\text{NR}}$, that is real far from resonances.  Therefore, CARS intensity does not scale linearly with molecular concentration and suffers from a strong non-resonant background (NRB), originating from four-wave mixing processes, which coherently mixes with the resonant CARS signal of interest. This NRB leads to distorted spectral line shapes and shifted vibrational peaks, complicating the interpretation of CARS spectra \cite{Rigneault2018}. Consequently, the measured CARS intensity does not directly reflect the spontaneous Raman response of the sample, limiting its chemical specificity.

Although point-scanning CRS techniques provide detailed spectroscopic information, their acquisition speed is limited by the scanning mirrors, and they are susceptible to motion artifacts and photodamage due to localized high intensities. Wide-field microscopy offers an attractive alternative, enabling camera-based detection and single-shot imaging over large fields of view. While wide-field implementations have been demonstrated for CARS \cite{Ceconello2022, Fantuzzi2023, Berto2012, Berto2013, Vernuccio2024}, they remain challenging for SRS due to the need for high dynamic range lock-in camera systems capable of demodulating at high repetition rates. As a result, SRS is typically restricted to point-scanning configurations, whereas wide-field approaches have primarily been explored for CARS. However, effective NRB suppression remains essential for extracting chemically meaningful information from CARS measurements.

Over the past decades, several strategies have been proposed to mitigate or remove the NRB. A first class of approaches relies on numerical reconstruction algorithms that retrieve the resonant susceptibility from intensity-based CARS spectra. Methods such as the time-domain Kramers--Kronig (TDKK) transform \cite{Camp2015, Cicerone2012, Liu2009} and the maximum entropy method (MEM) \cite{Vartiainen1992} enable reconstruction of Raman-like spectra from CARS measurements. More recently, deep-learning approaches \cite{Vernuccio2022artificial} have been introduced for NRB removal and spectral retrieval, including convolutional neural networks \cite{Valensise2020, Wang2022}, recurrent neural networks such as LSTM \cite{Houhou2020} and Bi-LSTM \cite{Junjuri2023} architectures, hybrid CNN-GRU models \cite{Vernuccio2024_NRB}, and generative adversarial networks \cite{Vernuccio2024_NRB}. While these approaches can be effective, they typically rely on training datasets, numerical assumptions, or prior knowledge of the spectral response.
An alternative strategy focuses on experimental suppression of the NRB. Techniques such as polarization-resolved CARS \cite{Cheng2001_polariz} and time-resolved CARS \cite{volkmer2002time, Hempel2025} detection schemes can reduce the non-resonant contribution. However, these methods typically require complex optical setups and often reduce the CARS resonant signal strength.
A more fundamental solution involves retrieving the complex anti-Stokes field, rather than only its intensity. Access to both amplitude and phase of the CARS field enables direct separation of the resonant and non-resonant contributions, since the vibrational information of interest is contained in the imaginary component of the nonlinear susceptibility, which is directly related to the spontaneous Raman signal. Heterodyne detection schemes \cite{Dudovich2002, Potma2006, Jurna2009} have been proposed to measure the CARS phase, in which the CARS signal interferes with a reference beam. While powerful, these approaches generally require interferometric stability and precise alignment of a reference field, which complicates their implementation, particularly in wide-field configurations.


Wavefront-imaging techniques offer an appealing alternative, as they enable direct wide-field measurement of both the amplitude and phase of an optical field without the need for a separate reference beam. Multiwave lateral shearing interferometry was introduced in the 90’s as a high-resolution wavefront sensing alternative to standard Shack-Hartmann wavefront sensor \cite{Primot:93, Primot:95}. The most advanced version, known as quadriwave lateral shearing interferometry (QLSI), is based on a Modified Hartmann Mask (MHM) \cite{Primot2000}, which consists of a 2D diffraction grating (also later called cross-grating) optimized to create four replicas only. Four replicas are required to retrieve the wavefront gradients in both orthogonal directions (x and y), which is necessary for complete two-dimensional phase reconstruction. In 2009, the technique was extended to quantitative phase imaging microscopy \cite{Bon:09}. Recently, QLSI received lots of attention for its broad quantitative phase imaging applicability, ranging from biosciences to nanophotonics \cite{Baffou:2012, Bon:2017,Baffou2023, Gentner2024,Wu2024, Nguyen2023, Aknoun2015}. QLSI is thus based on two simple elements: a camera and a 2D grating, separated by less than 1 mm from each other. The transmittance complex amplitude of the 2D grating is a three-level approximation of a perfect sine in directions x and y. It features the superposition of two gratings: one absorbing (vertical and horizontal lines) and one phase only (chessboard pattern with 0 and $\pi$ phase shift) \cite{Primot2000}. When the light impinges on the grating, it splits into four laterally sheared replicas that interfere - after limited propagation - at the camera sensor (see Figure \ref{Figure 1}(a)). Any local curvature in the impinging wavefront ($W$) creates distortions on the interferogram. The analysis of the interferogram enables the retrieval of information about the intensity and gradient of the wavefront \cite{Bon:09, Bon:2017, Baffou2023, Baffou2021}. Integration of this gradient yields the relative optical path difference (OPD) $\delta L = W$, which can be converted into phase measurements for a monochromatic beam with wavelength $\lambda$ through the following relationship:
\begin{equation}
    \phi = \frac{2\pi}{\lambda} W
    \label{equation_phase}
\end{equation}

\textcolor{black}{Early studies demonstrated the feasibility of using wavefront imaging to capture the CARS phase \cite{Berto2012,Berto2013}. However, these pioneering implementations, which utilized plane-wave illumination, reported very small fields of view ($<$5$\text{\textmu m}$), required long integration times (tens of seconds), and exhibited very slow spectral tunability. To suppress the NRB, one of these approaches applied polarization-resolved detection \cite{Berto2012}. While effective, this strategy inevitably attenuates the resonant signal of interest and distorts the Raman spectrum, as the relative intensity of each vibrational mode depends on its depolarization ratio \cite{Munhoz2011}. In contrast, the second approach \cite{Berto2013} employed a complex illumination scheme based on a highly folded boxCARS geometry to reduce the axial extent of the excitation volume. This configuration distorts the sample image and generates an interference pattern in the final reconstruction.}

In this work, we demonstrate a phase-resolved wide-field hyperspectral CARS microscopy technique that enables direct retrieval of the complex anti-Stokes field over a field of view significantly larger than previously reported implementations (FoV = 60 $\times$ 60 $\text{\textmu m}^2$) at a frame rate of 1.4 Hz without using polarization CARS schemes or complex illumination geometries. \textcolor{black}{Our approach relies on the combination of a laser source with high-energy ($\text{\textmu J}$-level) and low-repetition-rate (200 kHz) picosecond pulses, random speckle illumination, and QLSI.  The powerful picosecond laser source enables activating the nonlinear CARS excitation over a large field-of-view. The speckle excitation serves two purposes. (i)} It distributes the optical intensity over a large area at the back pupil plane of the objective, thereby preventing optical damage while maintaining sufficiently strong nonlinear excitation and enabling imaging over a large field of view. \textcolor{black}{(ii)} It suppresses coherent imaging artifacts that typically arise in wide-field nonlinear microscopy, resulting in artifact-free image formation. \textcolor{black} {QLSI, integrated into a wide-field CARS system, simultaneously measures the amplitude and phase and estimates the complex CARS field of the sample}. This strategy extracts the vibrational phase to separate the resonant contribution from the non-resonant one. From this isolated resonant signal, we reconstruct an NRB-free hyperspectral wide-field CARS image.
We demonstrate this capability through wide-field phase-resolved CARS imaging of complex microplastics and biological samples, retrieving vibrational spectra across the CH-stretching region (2800-3100 cm$^{-1}$).

\section{Theory}
Coherent Anti-Stokes Raman Scattering (CARS) is a third-order nonlinear optical process used to probe the vibrational modes of a molecule. It relies on the interaction of multiple laser fields with a medium to generate a coherent signal at a new frequency. Although comprehensive tutorials cover the general background \cite{polli_2018,Rigneault2018,CARS_book}, we review only the principles required to understand a phase-sensitive CARS measurement.

The CARS process involves three optical fields, two pump fields at frequency $\omega_p$ and one Stokes field at frequency $\omega_s$. When the frequency difference matches the molecular vibrational resonance, i.e., $\omega_p - \omega_s = \Omega$, the resulting coherent molecular motion scatters the incoming fields spectrally, providing the anti-Stokes field at frequency $\omega_{as} = 2\omega_p - \omega_s$.

The process can be understood as a four-wave mixing mediated by the third-order nonlinear susceptibility $\chi^{(3)}$ of the medium. The induced nonlinear polarization due to the interaction of the two pump fields $\bm{E}_p$ and the Stokes field $\bm{E}_s$ with the sample can be written as \cite{Rigneault2018}:

\begin{equation}
\bm{P}^{(3)}(t) \propto \chi^{(3)} \bm{E}_p^2 \bm{E}_s^*,
\end{equation}

This nonlinear polarization radiates the CARS field amplitude $E_{as}=A_{as}e^{i(k_{as}z-\omega_{as}t)}+c.c.$, where $A_{as}$ denotes the complex envelope. Under the slowly varying envelope approximation (SVEA), $A_{as}$ satisfies:

\begin{equation}
    \frac{\partial A_{as}}{\partial z} \propto P_{as} e^{(i\Delta k z)}
\end{equation}

with $\Delta k = 2k_p-k_s - k_{as}$ the wave-vector mismatch. From the previous equation, assuming propagation in a medium with length $L$:

\begin{equation}
A_{as}(L) = \int_0^L \frac{\partial A_{as}}{\partial z} \, dz
\propto \chi^{(3)}(\omega_{as}) \, L \,
\mathrm{sinc}\!\left(\frac{\Delta k L}{2}\right)
e^{i \frac{\Delta k L}{2}} \, A_p^2 A_s\textcolor{black}{^*}
\label{equation_field_CARS}
\end{equation}

The anti-Stokes intensity is proportional to the square modulus of the generated field:
\begin{equation}
I_{as}(L) \propto \left| A_{as}(L) \right|^2\propto\left| \chi^{(3)}(\omega_{as}) \right|^2
L^2 \, \mathrm{sinc}^2\!\left(\frac{\Delta k L}{2}\right)
\, I_p^2I_s
\end{equation}
where $I_p \propto |A_p|^2$ and $I_s \propto |A_s|^2$.

The susceptibility $\chi^{(3)}$ contains both a resonant contribution 'R' from molecular vibrations and a non-resonant contribution 'NR' from the instantaneous electronic response of the sample:
\begin{equation}
\chi^{(3)} = \chi^{(3)}_{\text{R}} + \chi^{(3)}_{\text{NR}} = \sum_{i=1}^{M} 
\frac{a_i}{\textcolor{black}{\delta} \textcolor{black}{-} i \Gamma_i} + \chi^{(3)}_{\text{NR}}
\end{equation}
\textcolor{black}{where $\delta = \omega_p -\omega_s - \Omega_i$ is the Raman detuning.}
The complex resonant term provides chemically specific information and can be modeled as a sum of Lorentzian peaks with amplitude $a_i$, centered at $\Omega_i$ and with \textcolor{black}{half}-width at half maximum $\Gamma_i$, while the non-resonant term is usually real and results from non-resonant electronic contributions. 
Because the anti-Stokes field superimposes resonant and non-resonant light, a conventional CARS detector captures only an intensity that scales with the squared magnitude of this combined field. Consequently, the resonant and non-resonant components interfere. This interference yields an independent intensity contribution for each component alongside a coupled cross-term. Together, these terms distort the recorded spectrum and obscure the pure Raman signature. The true spontaneous Raman spectrum resides exclusively within the imaginary part of the resonant third-order nonlinear susceptibility $\chi^{(3)}$ \cite{Hellwarth_77}. To extract this precise information from a standard CARS measurement, researchers have developed complex retrieval algorithms \cite{Camp2015, Cicerone2012, Liu2009,Vartiainen1992}. QLSI bypasses this entire complication and directly captures the complex-valued anti-Stokes field, isolates the imaginary component, and instantly gives access the uncorrupted Raman spectral data.

\begin{figure}[tp]
    \centering
    \includegraphics[width=\textwidth]{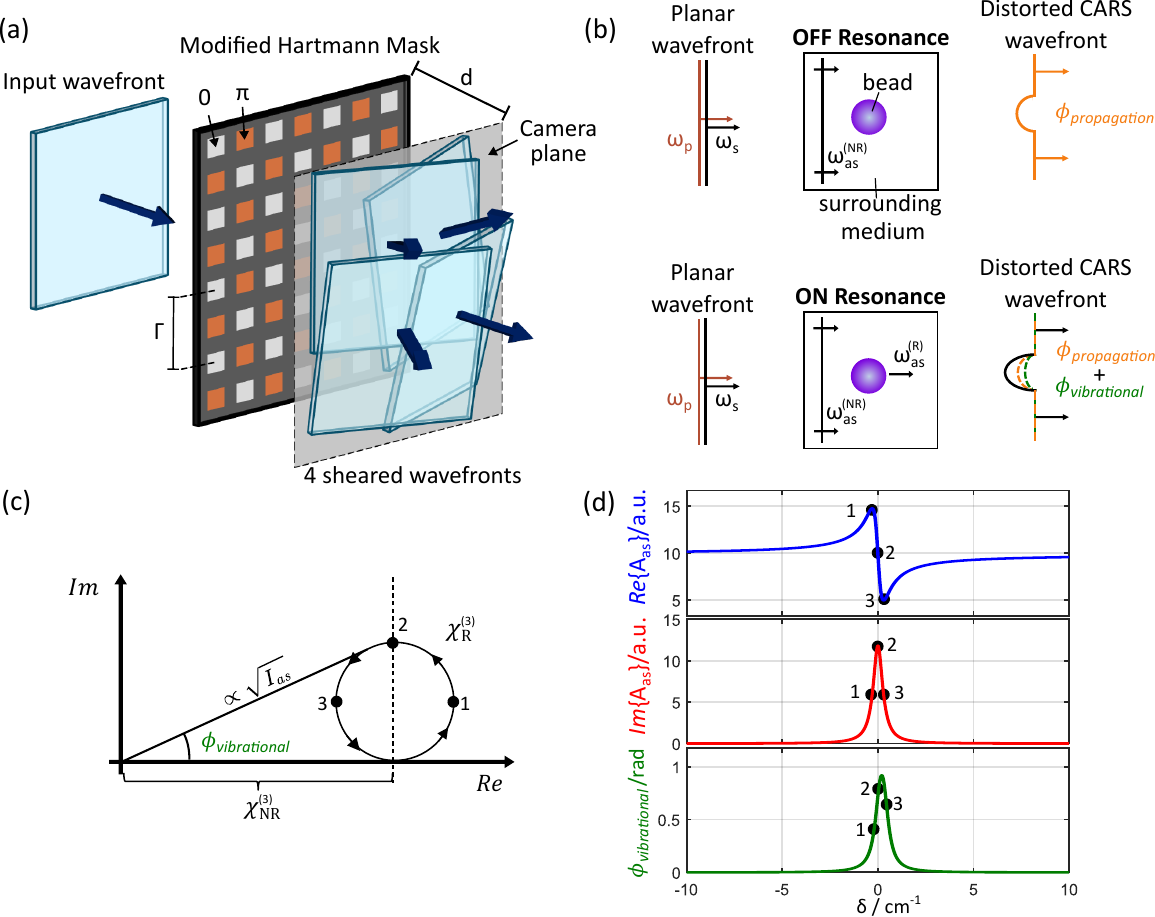}
    \caption{(a) Schematic of the quadri-wave lateral shearing interferometry  (QLSI) setup, including the 2D diffraction grating (modified Hartmann mask / cross-grating) and the camera with the 4 lateral sheared replicas of the input wavefront. (b) Principle of wavefront imaging in wide-field CARS microscopy: at OFF resonance conditions, the CARS wavefront is distorted only by the propagation phase, while at ON resonance conditions, the CARS wavefront is distorted by the propagation phase and an extra-phase shift, namely the vibrational phase.  (c) The third-order nonlinear susceptibility $\chi^{(3)}$ for a single Raman resonance can be represented in the complex plane as a circle (Lorentzian peak) shifted on the real axis by the nonresonant term of the nonlinear susceptibility. (d) From the graph in (c) one can extract the value of the real and imaginary part of the CARS electric field and the vibrational phase \cite{Rigneault2018}.}
    \label{Figure 1}
\end{figure}

To visualize these mechanics, we map the complex nonlinear susceptibility $\chi^{(3)}$ in the complex plane (see Figure \ref{Figure 1}(c)). A vector along the real axis represents the purely real, non-resonant component. In contrast, a classical simple harmonic oscillator models the resonant component. When the laser frequency sweeps across the resonance, the tip of this resonant vector traces a perfect circle. Spectral detuning alters both the length and the phase angle of this vector. To find the total susceptibility, we sum these two components vectorially. This addition shifts the entire resonant circle along the real axis. 

The angle $\phi_{\mathrm{vibrational}}$ of this total vector dictates the phase of the anti-Stokes field (Figure \ref{Figure 1}(c)). Consequently, this phase value relies on the magnitudes of both vectors alongside the intrinsic phase of the resonant component. While each localized point source in the medium contributes an individual response, the macroscopic anti-Stokes field emerges from the coherent superposition of all individual emissions at the sensor. To evaluate this collective signal, we propagate and sum the fields from every source through a Green's function. Equation \ref{equation_field_CARS} describes this spatial transport for a plane wave under the approximation of a slowly modulated susceptibility. From this geometric representation, we readily isolate the vibrational phase $\phi_{\mathrm{vibrational}}$ of the CARS field (see plot in Figure \ref{Figure 1}(d)), while its amplitude is given by $|A_{as}| = \sqrt{I_{as}}$. The corresponding real and imaginary parts of the anti-Stokes field (see Figure \ref{Figure 1}(d)) can be calculated as:

\begin{align}
\Re\{A_{as}\} &= \sqrt{I_{as}} \cos(\phi_{\mathrm{vibrational}}) \label{equation_real_part} \\
\Im\{A_{as}\} &= \sqrt{I_{as}} \sin(\phi_{\mathrm{vibrational}}) \label{equation_imaginary_part}
\end{align}

QLSI usually measures the phase shift accumulated by light propagating through the sample in transmission. The estimation of the phase of the CARS field with QLSI is based on the measurement of the anti-Stokes field with the sample and the measurement of a reference anti-Stokes field generated by identical pump and Stokes fields in a uniform background medium. In the data analysis, the sample-generated field is compared directly with this uniform-background reference field. The measurement, therefore, estimates the phase difference between two generated fields (the sample and the background) rather than an absolute anti-Stokes phase. This phase difference is the relevant quantity for characterizing the CARS field generated by an object embedded in a background medium.

The expression of the complex sample and reference fields at the QLSI grating-plane coordinate $\mathbf{r}_g$ and Raman detuning $\delta$ can be written as:

\begin{equation}
U_{\mathrm{samp}}(\mathbf{r}_g,\delta)
=
\left|U_{\mathrm{samp}}(\mathbf{r}_g,\delta)\right|
e^{i\Phi_{\mathrm{samp}}(\mathbf{r}_g,\delta)},
\end{equation}

\begin{equation}
U_{\mathrm{ref}}(\mathbf{r}_g,\delta)
=
\left|U_{\mathrm{ref}}(\mathbf{r}_g,\delta)\right|
e^{i\Phi_{\mathrm{ref}}(\mathbf{r}_g,\delta)}.
\end{equation}

Here, $\mathbf{r}_g$ denotes the field plane immediately incident on the grating.

\textcolor{black}{
Using equation 4 that describes the amplitude of the anti-Stokes component in a medium with length $L$, $U_{\mathrm{samp}}(\delta)$ and $U_{\mathrm{ref}}(\delta)$ can be expressed as follows:}
\begin{equation}
U_{\mathrm{samp}}(\delta) = \kappa_{\mathrm{samp}}(\delta)\,\chi^{(3)}_{\mathrm{samp}}(\delta)\,A_p^2A_s^{*}
\end{equation}
\begin{equation}
U_{\mathrm{ref}}(\delta) = \kappa_{\mathrm{ref}}(\delta)\,\chi^{(3)}_{\text{bg}}\,A_p^2A_s^{*}
\end{equation}

\textcolor{black}{where $\kappa(\delta)$ collects the interaction length, phase-matching sinc, and wavevector-mismatch phase of Eq. 4 for each arm — quantities set by the propagation properties of the medium. $\chi^{(3)}_{\text{samp}}$ denotes the nonlinear susceptibility of the sample, while $\chi_{\text{bg}}^{(3)}$ is the susceptibility of the background non-resonant medium.}

The normalized field ratio therefore can be written as:
\begin{equation}
U_{\text{norm}}(\delta) = \frac{U_{\mathrm{samp}}(\delta)}{U_{\mathrm{ref}}(\delta)} = \alpha(\delta)\,\frac{\chi^{(3)}_{\mathrm{samp}}(\delta)}{\chi^{(3)}_{\text{bg}}}
\end{equation}
and $\alpha(\delta) \equiv \kappa_{\mathrm{samp}}(\delta)/\kappa_{\mathrm{ref}}(\delta) = |\alpha(\delta)|\,e^{i\phi_{\text{propagation}}(\delta)}$.

Since $U_{\text{norm}}$ is the product of $\alpha$ and the complex susceptibility ratio, its phase \textcolor{black}{ $\phi(\delta)$ at every position (x,y)} is the sum of their respective phase:
\begin{equation}
\phi(\delta) = \arg \{U_{\text{norm}}(\delta)\} = \phi_{\text{propagation}}(\delta) + \phi_{\text{vibrational}}(\delta)
\end{equation}
with $\phi_{\text{vibrational}}(\delta) \equiv \arg \{\chi^{(3)}_{\mathrm{samp}}(\delta)$\}. Therefore, the phase of the normalized CARS field encompasses the phase due to propagation mismatch between the object and the background, $\phi_{\text{propagation}}$, that is present both on and off resonance, and an additional vibrational phase, $\phi_{\text{vibrational}}$, that adds up near resonance as illustrated in Figure \ref{Figure 1}(b).

By definition,
\begin{equation}
\sin(\phi_{\text{vibrational}}) = \frac{\Im\{\chi^{(3)}_{\text{samp}}\}}{|\chi^{(3)}_{\text{samp}}|} = \frac{\Im\{\chi^{(3)}_{\text{samp,R}}\}}{|\chi^{(3)}_{\text{samp}}|}
\end{equation}
since $\chi^{(3)}_{\text{samp,NR}}$ is assumed to be real.

Although $|\chi^{(3)}_{\text{samp}}|$ is not measured directly, it is derived from $|U_{\text{norm}}(\delta)|$ that is obtained experimentally. $|U_{\text{norm}}| = |\alpha||\chi^{(3)}_{\text{samp}}|/\chi^{(3)}_{\text{bg}}$, so $|\chi^{(3)}_{\text{samp}}| = \chi^{(3)}_{\text{bg}}\,|U_{\text{norm}}(\delta)|/|\alpha|$. Therefore, we can retrieve
\begin{equation}
\begin{aligned}
\Im\{\chi^{(3)}_{\text{samp,R}}(\delta)\}
&= \frac{\chi^{(3)}_{\text{bg}}}{|\alpha|}\,|U_{\text{norm}}(\delta)|
\sin\bigl(\phi_{\text{vibrational}}(\delta)\bigr) \\
&= \frac{\chi^{(3)}_{\text{bg}}}{|\alpha|}\,|U_{\text{norm}}(\delta)|
\sin\bigl(\phi(\delta)-\phi_{\text{propagation}}(\delta)\bigr)
\end{aligned}
\end{equation}
\textcolor{black}{Importantly, this expression remains valid for any $\chi^{(3)}_{\mathrm{samp}}(\delta)/\chi^{(3)}_{\text{bg}}$ ratio. 
}
In Section~\ref{dataacquisition}, we explain how $\phi_{\text{vibrational}}$ is extracted from hyperspectral CARS data, notably how the propagation phase is taken into account by baseline de-trending.

\section{Methods}
\subsection{Experimental setup}
 \begin{figure}[tp]
    \centering
    \includegraphics[width=\textwidth]{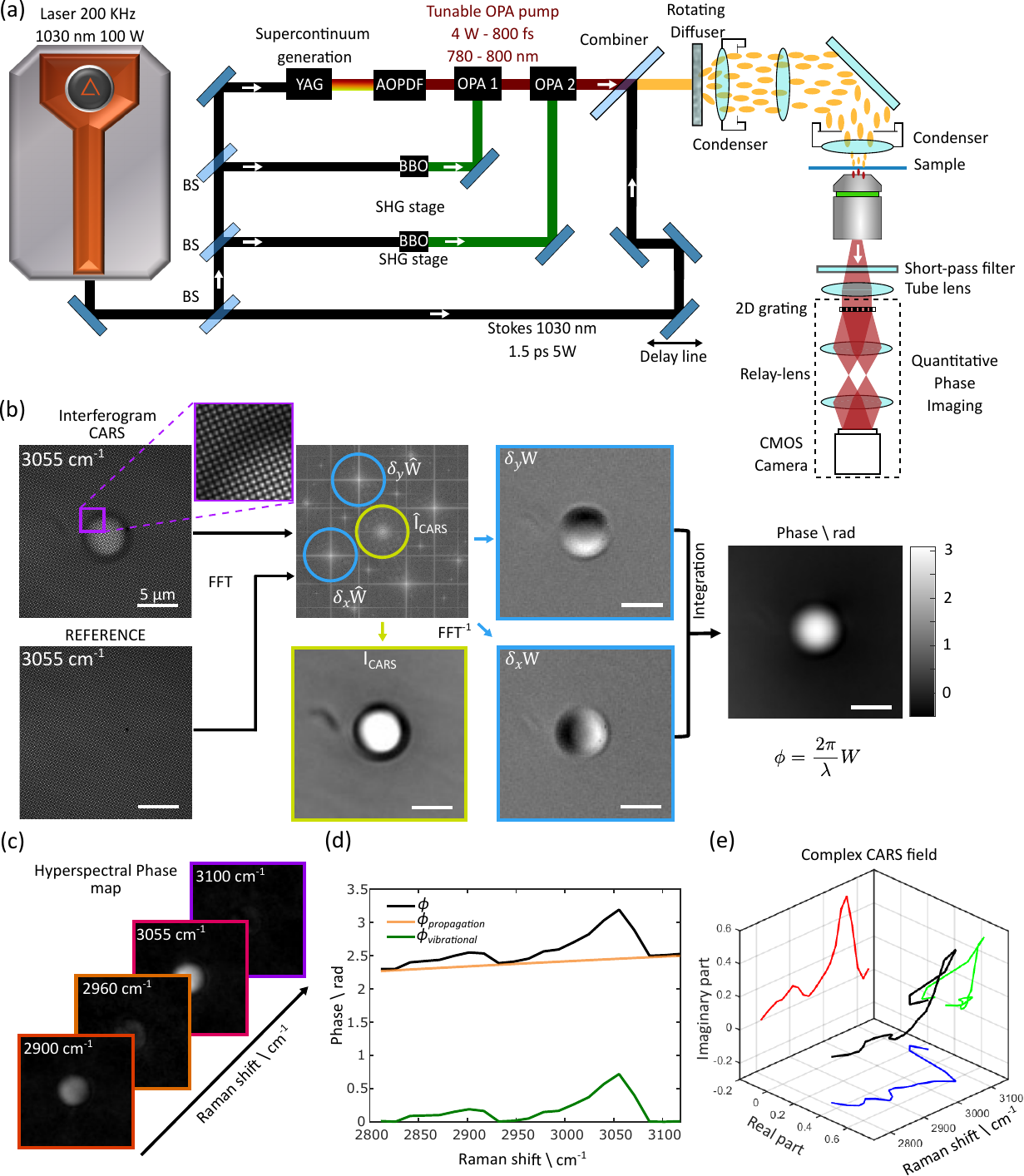}
    \caption{(a) Schematic of the experimental setup. BS: beam splitter; SHG: second harmonic generation; AOPDF: Acousto-optical programmable dispersive filter; OPA: optical parametric amplifier. (b) QLSI CARS interferogram image of a 5-$\text{\textmu m}$ polystyrene (PS) bead in water and its processing to extract the phase map for each Raman shift. (c) Hyperspectral vibrational phase map at different Raman shifts. (d) The measured PS bead spectral phase of a single pixel contains both a propagation $\phi_{\text{propagation}}$ and a vibrational $\phi_{\text{vibrational}}$ phase contribution. $\phi_{\text{vibrational}}$ is retrieved by removing $\phi_{\text{propagation}}$ through baseline detrending (orange line). (e) 3D-representation of the complex CARS field for the PS bead. Each resonance corresponds to a loop in the complex plane.}
    \label{Figure 2}
\end{figure}

Our experimental setup is based on a 100 W, 200 kHz, 1030~nm Yb laser (Tangor, Amplitude Laser), which serves as the Stokes beam and pumps a rapidly tunable picosecond optical parametric amplifier (OPA, Fastlite) covering the 700-900~nm wavelength range (see Figure \ref{Figure 2}(a)) and that is used as the pump beam. \textcolor{black}{Such a combination of pump and Stokes pulses enables a spectral resolution of 22 cm$^{-1}$.} The details of this laser system can be found in \cite{Vernuccio2024}, where an acousto-optic programmable dispersive filter (AOPDF) is used for fast (kHz) wavelength tuning. The pump and Stokes beams are temporally overlapped by means of a mechanical delay line placed on the Stokes path and spatially combined by a dichroic mirror and redirected onto a rotating diffuser (5°
 divergence, Edmund Optics) to generate random speckle patterns.
The diffuser plane is conjugated by means of a condenser lens (Nikon Abbe NA = 0.90) and a 200 mm lens to the back pupil plane of the illumination condenser lens (Nikon Abbe NA = 0.90). An iris is placed at the back pupil plane of the illumination condenser lens to tune its effective NA and directly control the incident angle of the speckle grains at the sample plane. A low illumination NA is preferred in QLSI to avoid reduced phase signal due to modulation transfer function drop \cite{Bon2014}. Moreover, the control of the NA enables fine-tuning the size of each speckle grain on the sample plane, without modifying the illuminated field of view. For our experiments, we fixed the illumination condenser NA to 0.4.
The wide-field anti-Stokes emission is collected in transmission using a 40$\times$/0.75~NA air objective (Nikon Plan Fluor NA = 0.75) and a 500 mm tube lens, for a total detection magnification of 100 between the sample and the QLSI camera plane. After removing the residual pump and Stokes beams with a low-pass filter (FESH0700, Thorlabs), the CARS signal is imaged onto a custom-built QLSI detection module coupled to a CMOS camera (OrcaFlash, Hamamatsu) (Figure~\ref{Figure 2}(a)). The rotating diffuser ensures that multiple distinct speckle patterns illuminate the sample during each camera exposure, and their temporal averaging produces homogeneous illumination. Speckle illumination not only prevents potential damage to the illumination condenser and the collection objective lenses but also suppresses imaging artifacts arising from the coherent nature of the CARS signal.


In our setup, we use a custom-made phase camera where the 2D diffraction grating is re-imaged to the camera sensor using a relay-lens system with a magnification of $Z$=1.11 \cite{Marthy_2019}. The 2D grating features a $\Gamma = $35.1 $\text{\textmu m}$ pitch and produces an interferogram at the camera plane with period $\frac{Z\Gamma}{2}$. Each interference fringe can be properly sampled using 3x3 camera pixels (Nyquist criterion, camera pixel size $p$ = 6.5 $\text{\textmu m}$). The grating image is positioned at a distance $d$ = 0.6 mm from the camera sensor. For calibration of the phase camera, a red LED (M660L4, Thorlabs) combined with a bandpass filter (BP643-20) was used to match the spectral range of the detected anti-Stokes field.

\subsection{Data acquisition and processing pipeline}
\label{dataacquisition}

The collected raw data consist of QLSI hyperspectral CARS images of the sample embedded in the surrounding medium (water) along with corresponding QLSI hyperspectral reference CARS images acquired over a homogeneous field of view containing only the surrounding medium (see the example for a 5 $\text{\textmu m}$ polystyrene (PS) bead in water in Figure~\ref{Figure 2}(b)). In QLSI, the OPD image of the reference is always subtracted from the OPD image of the sample to remove incoming wavefront phase distortions that are not associated with the sample but with both the wavefront distortions of the illuminating beam and the optical system aberrations \cite{Bon:09}. During acquisition, the tunable OPA, serving as the pump source, is tuned over the wavelength range 780–800 nm, while the Stokes beam is fixed at 1030 nm. This choice of wavelengths enables probing the entire CH-stretching region (2800 - 3100 cm$^{-1}$) of the vibrational spectrum of the sample.

\textcolor{black}{Data acquisition is performed by alternating the sample position between the CARS (containing the sample) and reference CARS (water) fields of view at each excitation wavelength. For each recorded image, an average of five consecutive frames is computed. The consecutive acquisition of sample and reference images, together with frame averaging, is adopted to optimize the phase reconstruction by minimizing the impact of laser instability and of slow temperature-dependent drifts in the system. During the averaging, we account for the full-frame readout time of the Hamamatsu ORCA-Fusion camera, which in standard scan mode is 42.99 ms. Moreover, the acquisition time for acquiring the full hyperspectral stack includes the time required for laser wavelength tuning (500 ms) and sample-stage translation between the sample and reference positions (1000 ms). We set these delays to ensure that the laser has stabilized at the selected wavelength and that the sample stage has completed its movement before image acquisition. Image saving is performed during the stage movement and therefore does not constitute an additional acquisition step. For a single-frame exposure time of 100ms, 5 averages per image, and 21 spectral points, which provide a sampling interval sufficient to identify the Raman peaks of the investigated chemical species, the entire hyperspectral dataset is acquired in approximately 70~s.}

Following acquisition, the QLSI CARS interferogram images are processed (using an algorithm adapted from standard QLSI \cite{Baffou2021, phaselab_baffou}) to retrieve, for each pair of CARS and reference CARS images at every Raman shift, the corresponding intensity and phase maps, the latter being obtained from the OPD map following Eq. \ref{equation_phase} (Figure~\ref{Figure 2}(b)). The phase maps are then corrected using custom-developed software to remove wavefront distortions via surface fitting. During this step, object regions are excluded to prevent bias in the baseline estimation.
An absolute phase reference is established by setting the phase to zero at a pixel of the hyperspectral phase map (Figure~\ref{Figure 2}(c)) corresponding to the surrounding medium, enabling quantitative phase retrieval of the sample. \textcolor{black}{ Moreover, this step enables accounting for common-mode phase distortions arising from laser and instrumental instabilities and aligns the phase maps by removing the arbitrary additive constant for each Raman shift coming from the integration step.}  Prior to extracting the real and imaginary components, a baseline detrending is applied independently to each pixel spectrum to remove spectral phase slopes arising from dispersion of the refractive index as well as thickness variations across the field of view that both affect the propagation phase $\phi_\text{propagation}$. The baseline detrending enables the extraction of the vibrational phase $\phi_\text{vibrational}$ (Figure \ref{Figure 2}(d)).

Finally, the real and imaginary parts of the complex CARS field are extracted. The imaginary part (Eq.\ref{equation_imaginary_part}) corresponds to the purely resonant vibrational response of the sample and can be directly compared to spontaneous Raman spectra, while the real part (Eq.\ref{equation_real_part}) contains the dispersive contribution. The real and imaginary parts of the complex CARS signal enable a 3D representation (see the example on the CARS field of PS in Figure \ref{Figure 2}(e)). As expected, each Lorentzian resonance traces a characteristic loop in the complex plane.
\textcolor{black}{On a standard computer, the complete processing pipeline for a hyperspectral CARS image stack containing 1000 $\times$ 1000 pixels and 21 spectral points requires less than 30 s, including QLSI phase reconstruction ($<$2 s), surface fitting of the full stack (5 s), pixel-wise spectral baseline detrending (12 s), and spectral unmixing (typically $<$10 s).}

\section{Results}

 \begin{figure}[tp]
    \centering
    \includegraphics[width=\textwidth]{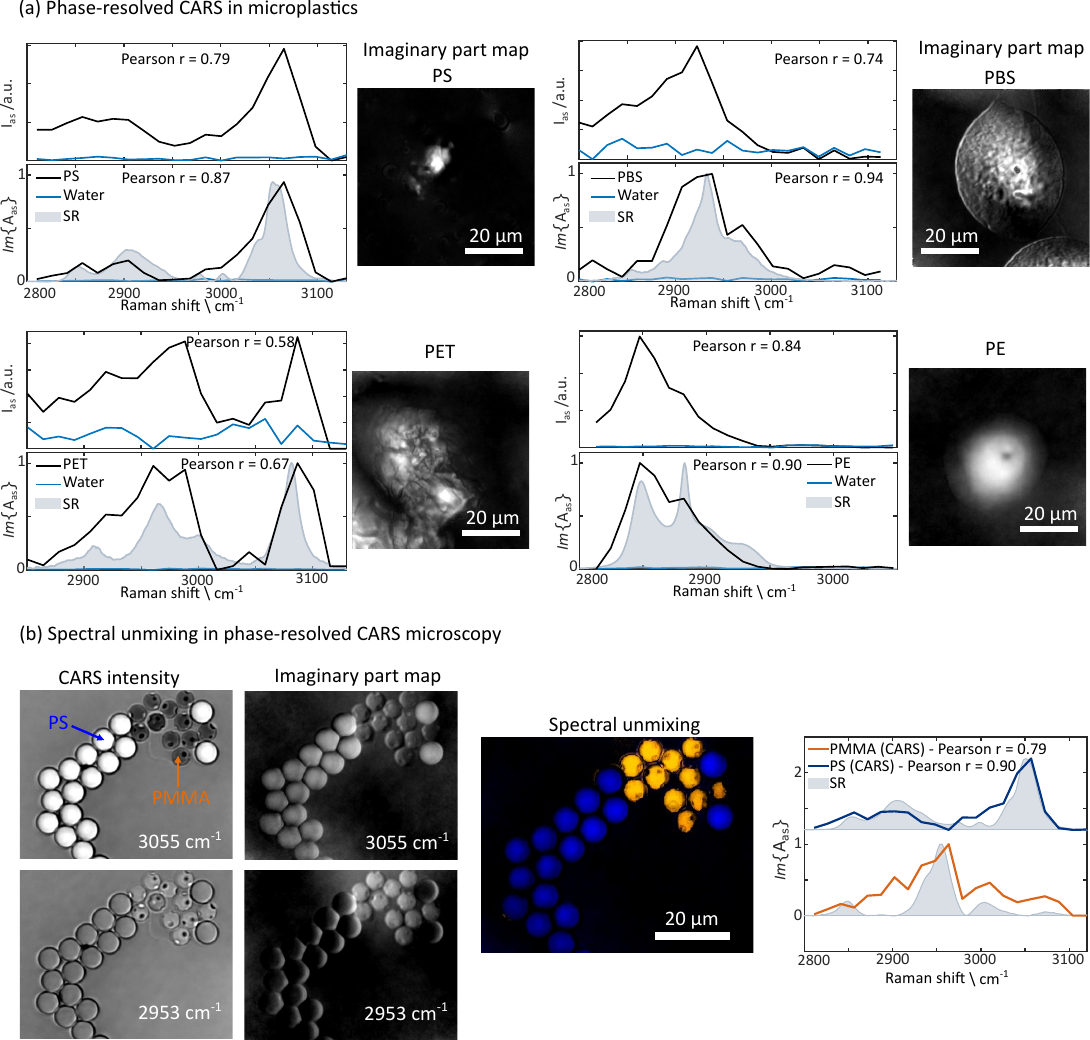}
    \caption{(a) Phase-resolved hyperspectral CARS microscopy of four different microplastics embedded in water: polystyrene (PS), polybutylene succinate (PBS), polyethylene terephthalate (PET), polyethylene (PE). For each sample, we represent the CARS intensity spectrum ($\text I_{as}$) and the reconstructed imaginary part ($\text{Im}\{A_{as}\}$) of the CARS field spectrum after extraction of the vibrational phase. The imaginary part of the CARS field (black line) can be directly compared to spontaneous Raman measurements (solid gray - SR). \textcolor{black}{The Pearson correlation coefficient (r) is reported for both the intensity and phase-resolved CARS spectra to quantify their similarity to the corresponding spontaneous Raman spectra.} Field of view: 60 $\times$ 60 $\text{\textmu m}^2$; total exposure time: 500 ms per Raman shift (5 averages of 100 ms); 21 spectral points. (b) Spectral unmixing in phase-resolved hyperspectral CARS microscopy for a sample of 6-$\text{\textmu m}$ polymethyl methacrylate (PMMA) beads and 7 $\text{\textmu m}$ polystyrene (PS) beads in index-matching medium (n = 1.53). The images show the two CARS intensity maps relative to the main peak in the CH-stretching region of PS (3055 cm$^{-1}$, top image) and PMMA (2953 cm$^{-1}$, bottom image), the imaginary part maps at the same Raman shifts after extraction of the vibrational phase, and the false-color image is obtained by spectral unmixing. The spectral unmixing is applied to the imaginary part map after NRB suppression, using the N-FINDR algorithm. The plot compares the endmembers given by the N-FINDR algorithm with the spontaneous Raman spectra of PS and PMMA, \textcolor{black}{together with the Pearson correlation coefficients (r)}. Field of view: 78 $\times$ 60 $\text{\textmu m}^2$; total exposure time: 2 s per Raman shift (5 averages of 400 ms); 21 spectral points.}
    \label{Figure 3}
\end{figure}

We demonstrated the imaging capability of the system by using samples of different types of microplastics in water that featured complex and non-symmetric structures. Measuring and identifying microplastics with CARS microscopy enables rapid, label-free chemical characterization with high spatial resolution, allowing reliable detection of particles in complex environmental and biological matrices. For each microplastic, we applied the previously described processing pipeline to derive, for each wavenumber, the imaginary part of the CARS field spectrum ($\operatorname{Im}\{A_{aS}\}$) which we compared with the spontaneous Raman spectrum. Spontaneous Raman spectra of microplastics were acquired using a commercial confocal Raman microspectroscopy system (LabRAM Soleil, Horiba Scientific) with 532 nm continuous-wave laser excitation. The Raman signal was collected in a backscattering configuration through a 20× objective (NA = 0.45) and analyzed using a 600 gr/mm diffraction grating. Spectra were recorded with a 1 s integration time and averaged over five accumulations. Background contributions were removed by polynomial baseline subtraction before analysis.
We performed measurements with the wide-field CARS system of four different types of microplastics diluted in water and sandwiched between two 170 $\text{\textmu m}$ glass coverslips: polystyrene (PS), polybutylene succinate (PBS), polyethylene terephthalate (PET), and polyethylene (PE) (Figure \ref{Figure 3}(a)). Images were acquired at a frame rate of \textcolor{black}{1.4 Hz per Raman shift (this value includes the full-frame readout time of our camera)}. While the CARS intensity spectra presents a distorted peak whose amplitude and position cannot directly be compared to spontaneous Raman data, the reconstructed imaginary part spectra match the spontaneous Raman one in position and relative amplitude of the peaks. \textcolor{black}{We quantified the similarity between the spontaneous Raman spectra and either the reconstructed imaginary part of the CARS field or the CARS intensity spectra using the Pearson correlation coefficient, $r$. For all measurements, the reconstructed imaginary part yielded a higher Pearson correlation coefficient than the corresponding CARS intensity spectrum, indicating a higher spectral similarity with the spontaneous Raman response.} We note that the distortion and shift of the peaks in the CARS intensity spectra are not similar for all the microplastics, indicating different strengths of the non-resonant contributions.

Once the system capabilities have been tested in collecting spectra of microplastics, we generated a mixture of two different microplastic beads of polymethyl methacrylate (PMMA) of 6 $\text{\textmu m}$ and polystyrene (PS) of 7 $\text{\textmu m}$ in an index-matching non-resonant medium (n = 1.53, Cargille Labs, Cargille Immersion Liquid). Figure \ref{Figure 3}(b) shows the CARS intensity and CARS field imaginary part image of the sample. \textcolor{black}{By applying a spectral unmixing algorithm based on N-FINDR \cite{Winter2004,Vernuccio2023} to the imaginary-part data, we identified three distinct endmembers corresponding to: water, which exhibits no Raman resonances, PS, and PMMA. These results demonstrate that the proposed phase-resolved wide-field CARS approach enables chemical imaging of heterogeneous samples while suppressing the non-resonant background (NRB).}

 \begin{figure}[tp]
    \centering
    \includegraphics[width=\textwidth]{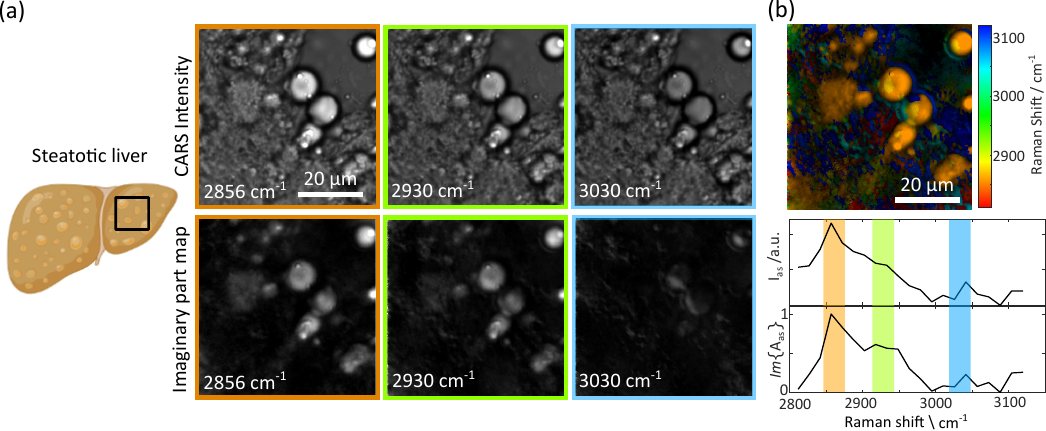}
    \caption{(a) Hyperspectral CARS intensity and imaginary CARS field images of a thin slice of a murine steatotic liver containing lipid droplets at three different Raman shifts: 2856 cm$^{-1}$ for CH$_2$ symmetric stretching, 2930 cm$^{-1}$ for CH$_2$ asymmetric and CH$_3$ symmetric stretching, and 3030 cm$^{-1}$ for the band of =C–H stretching vibrations. (b) HSV (Hue, Saturation, Value) representation of the retrieved NRB-free CARS field imaginary maps. The plot shows the characteristic intensity and retrieved NRB-free CARS spectrum of a lipid droplet in the images in panel (a). The colored rectangles indicate the Raman shifts corresponding to the frames in panel (a). Field of view: 60 $\times$ 60 $\text{\textmu m}^2$; total exposure time: 2 s per Raman shift (5 averages of 400 ms); 21 spectral points.}
    \label{Figure 4}
\end{figure}

Eventually, we tested our microscope on heterogeneous biological samples. We measured a 10 $\text{\textmu m}$ thin slice of a murine steatotic liver in water sandwiched between two 170 $\text{\textmu m}$ glass coverslips. In this context, CARS microscopy is a powerful imaging technique that rapidly provides, in a label-free way, precise localization and chemical identification of lipid droplets, enabling detailed insight into lipid accumulation. \textcolor{black}{In Figure \ref{Figure 4}(a), we showcase the CARS intensity and CARS field imaginary maps of the steatotic liver sample at three characteristic Raman shifts associated with lipid vibrational modes: 2856 cm$^{-1}$, corresponding predominantly to CH$_2$ (methylene group) symmetric stretching, 2930 cm$^{-1}$, associated with CH$_2$ asymmetric stretching and chain-end CH$_3$ (methyl group) symmetric stretching, and 3030 cm$^{-1}$, corresponding to =C–H stretching vibrations \cite{Talari2014, Czamara2014}. The removal of the NRB in the CARS field imaginary-part images enhances the chemical contrast and highlights lipid-rich structures.} \textcolor{black}{As in the case of microplastics, the reconstructed ($\operatorname{Im}\{A_{aS}\}$) spectrum (Figure \ref{Figure 4}(b)) is free of the NRB and recovers characteristic lipid-associated Raman features, such as the bands around 2860 and 2930 cm$^{-1}$ \cite{Czamara2014}. In Figure \ref{Figure 4}(b), we show an HSV (Hue, Saturation, Value) representation of the retrieved NRB-free CARS field imaginary maps, where the hue encodes the dominant Raman shift, the saturation is a constant (equal to 1), while the brightness (value) represents the corresponding signal intensity. This representation enables the identification of regions with a higher lipid abundance and highlights their spatial distribution within the tissue, providing a direct visualization of the chemical heterogeneity of the sample.}

\begin{figure}[tp]
    \centering
    \includegraphics[width=\textwidth]{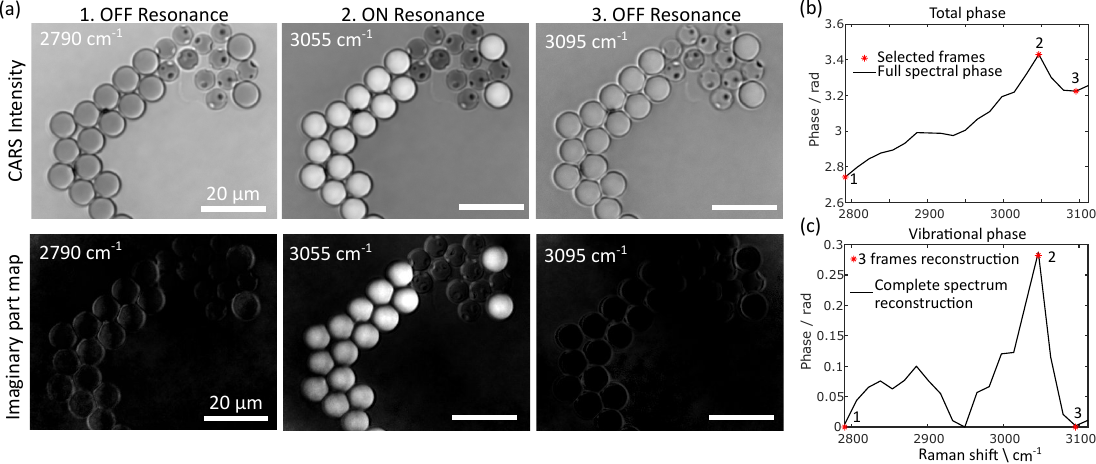}
        \caption{Reconstruction of the vibrational phase using 2 frames OFF resonance and 1 frame ON resonance. Sample of 6-$\text{\textmu m}$ polymethyl methacrylate (PMMA) beads and 7 $\text{\textmu m}$ polystyrene (PS) beads in index-matching medium (n = 1.53).  (a) CARS intensity maps at three different Raman shifts (2790 cm$^{-1}$ OFF Resonance, 3055 cm$^{-1}$ ON Resonance PS bead, and 3095 cm$^{-1}$ OFF Resonance) and reconstructed CARS field imaginary part maps after retrieval of the vibrational phase using only three frames. (b) The total spectral phase (propagation phase + vibrational phase) in a pixel of a PS bead using the full spectral information (black solid curve) and in correspondence with the three selected frames (red stars). (c) Vibrational phase reconstructed from the total spectrum with baseline detrending (black solid curve) and obtained using only three Raman shifts (red stars). }
    \label{Figure S1}
\end{figure}

\textcolor{black}{Notably, our proposed phase-resolved wide-field CARS approach offers a more practical route to extract the purely resonant component of the CARS field than numerical \cite{Camp2015, Cicerone2012, Liu2009, Vartiainen1992} or deep-learning algorithms \cite{Vernuccio2022artificial, Vernuccio2024_NRB}, which typically require densely sampled and broadband spectral data to accurately remove the NRB and reconstruct the resonant response. Because the refractive index variation with the scanned wavelength is approximately linear across the investigated spectral region (as shown in Figure \ref{Figure 2}(d)), the entire spectrum is not required to isolate the imaginary part of the CARS field. Instead, our algorithm requires a minimum of three pairs of images, each consisting of a CARS and a reference-CARS image acquired at the same Raman shift. One on-resonance pair is used to capture the combined vibrational and propagation contributions. Two off-resonance pairs are then used to characterize the linear variation of the propagation phase across the scanned wavelengths, enabling retrieval of the purely propagation component, $\phi_{\text{propagation}}$. By subtracting the interpolated propagation contribution from the total reconstructed phase, we isolate the purely vibrational phase $\phi_{\text{vibrational}}$.} \textcolor{black}{Figure \ref{Figure S1} illustrates this possibility of retrieving wide-field NRB-free chemical maps with only three Raman shifts — for this we have used the same dataset as in Figure \ref{Figure 3}(b). The sample contains polystyrene (PS) and polymethyl methacrylate (PMMA) beads in a non-resonant index-matching medium (n = 1.53). Figure \ref{Figure S1}(a) shows three CARS intensity maps corresponding to three Raman shifts for the PS Raman spectrum. The maps at 2790 cm$^{-1}$ (frame 1) and 3095 cm$^{-1}$ (frame 3) refer to two regions of the vibrational spectrum of PS with no resonances (off-resonance condition), while the map at 3055 cm$^{-1}$ (frame 2) corresponds to the main peak of PS in the CH-stretching region (on-resonance condition). We can estimate the propagation phase $\phi_\text{propagation}$ using the two off-resonance frames. We linearly interpolate it at the on-resonance Raman shift and subtract it from the total phase (in Figure~\ref{Figure S1}(b)) measured in frame 2. This procedure directly yields the purely vibrational phase. As shown in Figure~\ref{Figure S1}(c), the resulting vibrational phase is in excellent agreement with that obtained from full hyperspectral baseline detrending. The reconstructed CARS field imaginary part images at the three selected Raman shifts are also presented in Figure~\ref{Figure S1}(a).}

\textcolor{black}{This streamlined procedure allows for the reconstruction of CARS field imaginary-part maps without the full spectral acquisition, significantly reducing measurement time for samples with known Raman features.}

\section{Discussion and Conclusion}

Our phase-resolved CARS method enables the acquisition of NRB-free spectra of chemical species without the need for external reference beams. Unlike traditional methods such as time-domain CARS or polarization-resolved CARS, which often reduce the resonant signal contribution, our approach preserves signal integrity while simplifying the optical setup. This is achieved using QLSI quantitative phase imaging, where interference occurs directly between the four sheared replicas generated by a 2D diffraction grating. Although this methodology does not provide the measurement of an absolute CARS phase, it enables us to retrieve the CARS phase relative to the surrounding medium. Nevertheless, absolute phase reconstruction can still be achieved through hyperspectral measurements, provided that the field of view contains a non-resonant region to define a phase zero. For example, in all the presented images, there is always an area where the sample of interest is absent, and the surrounding medium is visible. We have demonstrated that the propagation phase can be decoupled from the vibrational phase by leveraging spectral information. Applying a baseline detrending algorithm effectively removes contributions from refractive-index mismatches and wavelength-dependent phase variations. 

\textcolor{black}{Furthermore, we showed a more practical route to extract the vibrational phase of a Raman resonance by using only three frames (Figure \ref{Figure S1}). Exploiting the quasi-linear dependence of the refractive index on wavelength over the investigated spectral range (see Figure \ref{Figure 2}(d)), NRB-free wide-field CARS maps can be reconstructed using only three pairs of images (three CARS and three reference-CARS frames) at three different Raman shifts. A single on-resonance image captures both vibrational and propagation contributions, while two off-resonance images are sufficient to characterize the linear spectral evolution of the propagation phase. By interpolating this propagation contribution and subtracting it from the total reconstructed phase, the purely vibrational phase is directly retrieved. Therefore, in wide-field CARS microscopy, QLSI provides spatially resolved phase information at each Raman shift without full spectral sampling. It substantially reduces acquisition time while preserving quantitative chemical specificity, making it particularly well-suited for samples with known or sparsely distributed Raman features. This approach differs fundamentally from intensity-based phase-retrieval methods such as TDKK \cite{Camp2015, Cicerone2012}. In TDKK, the phase is not directly measured but reconstructed from the intensity spectrum through the Kramers–Kronig relation, and accurate retrieval generally benefits from densely sampled and broadband spectral data. More precisely, the phase is reconstructed by imposing a relation between the log-amplitude and phase. This reconstruction is nonlinear, and a smooth mismatch between the NRB reference spectrum (acquired on a non-resonant medium and used to normalize the CARS spectra) and the true non-resonant susceptibility of the sample is converted, through this nonlinearity, into amplitude and phase errors. Spectral baseline detrending corrects the phase error, while the use of the Kramers-Kronig relation on the estimated phase error and unity-centering of the real component of the phase-corrected spectrum account for the amplitude error. In QLSI, instead, the phase is retrieved from the spatial wavefront at each Raman shift.}

\textcolor{black}{While our preliminary work demonstrated background-free QLSI-CARS \cite{Berto2012}, it remained impractical due to a small field of view ($5 \times 5\,\text{\textmu m}^2$), slow acquisition times ($40\,\text{s}$), and extremely slow ($>$30 minutes) hyperspectral  acquisitions. Here, we significantly advance the QLSI-CARS non-resonant background (NRB) removal concept by using a $\text{\textmu}\text{J}$-level, low-repetition-rate ($200\,\text{kHz}$) picosecond OPA. This enables $60 \times 60\,\text{\textmu m}^2$ field-of-view CARS phase images at $1.4\,\text{frames/s}$ across the entire CH-stretching region, rendering the technique suitable for practical applications such as microplastics identification and liver steatosis visualization.}

\textcolor{black}{These advances are also enabled by the use} of speckle illumination. Traditional wide-field schemes use collimated beams, which often result in high power densities and focal hotspots within the illumination and collection objectives back-focal planes, posing a risk of optical damage \textcolor{black}{(see estimation of the peak power density damage threshold in the supplemental document of \cite{Vernuccio2026})}. In contrast, the stochastic nature of the speckle pattern redistributes the intensity over a larger area at the back pupil plane of the objectives, preventing such damage. This advantage allows for the application of higher incident powers \textcolor{black}{enabling nonlinear imaging over large fields-of-view.}
Furthermore, speckle illumination ensures a more homogeneous excitation profile. By effectively scrambling the spatial coherence of the source, it suppresses interference artifacts, such as halos or fringes, that typically arise from the highly coherent nature of the CARS signal, resulting in cleaner and more accurate chemical maps. \textcolor{black}{We verified this effect by comparing spatially coherent collimated-beam illumination with speckle illumination generated using a rotating diffuser (data not shown). Consistent with these observations, previous studies have shown that a reduced temporal coherence of the laser source in QLSI can suppress coherent artifacts, such as speckle noise, and improve the signal-to-noise ratio of the reconstructed phase \cite{Ta2026}.}

\textcolor{black}{Our QLSI-CARS implementation is currently performed in transmission, and its extension to an epi-configuration is challenging. In epi-CARS, the detected signal is generally dominated by the forward-emitted CARS field that is back-scattered by sample inhomogeneities and interfaces, thereby introducing additional phase terms. If the spectral dispersion of these additional phase contributions can be detrended in the same manner as $\phi_{\text{propagation}}$, the proposed QLSI-CARS approach could, in principle, be extended to epi-detection. However, a dedicated study would be required to assess the feasibility of this approach and determine the extent to which the additional phase contributions can be reliably characterized and removed.}

To conclude, this work proposes a practical methodology to circumvent the inherent difficulties of performing background-free coherent Raman imaging. While SRS provides direct access to the imaginary, purely resonant component of the third-order nonlinear susceptibility, it requires a sophisticated demodulation detection scheme. However, translating SRS to a wide-field scheme is challenging and has not been demonstrated yet, mainly because it necessitates lock-in cameras with demodulation capability at high frequency (MHz). Conversely, our method uses CARS that can be implemented in wide-field but carries a non-resonant background that hampers its chemical imaging selectivity and its detection limit. Our phase-sensitive CARS provides a practical solution to remove the NRB in CARS wide-field imaging by retrieving the imaginary part of the CARS field, the latter being proportional to the imaginary part of $\chi^{(3)}_{\text{sample}}$, i.e. the spontaneous Raman signal. The proposed phase CARS measurement is straightforward to implement, requires only a rotating diffuser, a QLSI camera that is commercially available, and a high-power laser source to drive the nonlinear contrast.

\section*{Funding}
We acknowledge the financial support from the Centre National de la Recherche Scientifique (CNRS), A*Midex (Grant No. ANR-11-IDEX-0001-02, AMX-19-IET-002, AMX-25-REC-COFI-UE-001), ANR (Grant Nos. ANR-10-INSB-04-01, ANR-11-INSB-0006, ANR-16-CONV-0001, and ANR-21-ESRS-0002 IDEC), and INSERM Grant No. 22CP139-00. Chan Zuckerberg Initiative (DAF 2024-337798). This project has received funding from Horizon Europe (Marie Skłodowska-Curie Postdoctoral Fellowship Grant No. 101148683 CHIMERA) and the European Research Council (ERC, SpeckleCARS, Grant No.101052911, sCiSsoRS Grant No.101124764). 

\section*{Acknowledgment}
F.V. acknowledges funding from Horizon Europe project CHIMERA (Grant No. 101148683). The authors acknowledge support from Georges Farkouh for providing help with the spontaneous Raman measurements and Enora Prado (ifremer Brest) for providing the microplastic samples.

\section*{Disclosures}
The authors declare no conflicts of interest.

\section*{Data Availability Statement}
All data and codes supporting the findings of this study are available from the corresponding authors upon reasonable request.


\bibliographystyle{unsrtnat}
\bibliography{references}

\end{document}